# Расчёт сечений T(³He, np)⁴He и (Ty⁻)(³He, npy⁻) ⁴He в динамике трёх тел

М.В. Егоров[1,2]

[1]*Физический факультет, Томский государственный университет, г.Томск*

[2]*Лаборатория теоретической физики, Объединённый институт ядерных исследований, г.Дубна*

*В работе проведены трёхчастичные расчёты сечений реакции синтеза ³He+T→n+p+⁴He с помощью решения связанных шести интегральных уравнений Фаддеева с кластерным представлением ядра мишени - тритона как связанной нейтрон-дейтронной системы. Короткодействующие взаимодействия в подсистемах пар частиц включали микроскопические np, nd, n⁴He, n³He матрицы рассеяния, а также феноменологические d³He модели, допускающие связывание упругого канала с каналом реакции d³He. Кратко обсуждается роль моделей d³He и кулоновского взаимодействия ³He и T ядер. Полученная трёхчастичная матрица реакции использована для оценки роли мюона в процессе ³He+(Ty⁻)→n+p+⁴He+y⁻ с атомом Ty⁻, где y⁻∈[e⁻,µ⁻]. Показано, что эффект роста сечения вызванный присутствием мюона в зоне реакций, продиктован фазовым множителем и сохраняется в первом Борновском приближении для точной четырёхчастичной матрицы рассеяния.*

***Ключевые слова****: малочастичная динамика, ядерный синтез, гелий-3, тритий, уравнения Фаддеева*

## Введение

Реакция синтеза ядер гелия-3 $2\,^3\text{He} \to 2p\,^4\text{He}$ замыкает Солнечный протон-протонный цикл, и является одной из немногих реакций, все продукты которой являются заряженными частицами. Использование ядер ³He в реакциях синтеза примечательно тем, что только на шестом шаге возможных ядерных реакций ⁴He+³He→⁷Be[+e⁻→$\nu_e$+⁷Li]+γ →⁷Li+³He→⁴He+⁶Li→³He+⁶Li→d+⁷Be в смеси ядер ³He образуются дейтроны, которые через канал $dd \to n\,^3\text{He}$ приводят к нейтронам. Таким образом, практически вся энергия этой цепочки ядерных реакций остаётся в зоне синтеза, поддерживая его. Однако, несмотря на подробное экспериментальное восстановление сечения $2\,^3\text{He} \to 2p\,^4\text{He}$ его теоретическое исследование сопряжено с трудностью учёта кулоновского взаимодействия как в начальном, так и конечном состоянии, а также необходимостью рассматривать механизм вылета сразу трёх частиц в непрерывный спектр энергий. Корректное формулирование кулоновского рассеяния трёх тел само по себе представляет фундаментальную проблему как в конфигурационном [1], так и в импульсном представлениях [2], и здесь необходимы дальнейшие исследования и бенчмарк тесты различных подходов, что выходит далеко за замки данной работы. Изотопическим аналогом реакции $2\,^3\text{He} \to 2p\,^4\text{He}$, лишённым кулоновского трёхтельного взаимодействия является процесс $T + \,^3\text{He} \to n + p + \,^4\text{He}$ или в компактной записи T(³He, np)⁴He. Обе реакции допускают несколько каналов, идущих за счёт сильного взаимодействия с промежуточными ядрами ⁵Li, ⁵He (в т.ч. в возбуждённых состояниях), и для обоих процессов применимо кластерное представление ядра-мишени благодаря которому синтез ядер T³He и ³He³He можно рассматривать как синтез d³He с вылетом в непрерывный спектр нейтрона и протона, соответственно. Схематически синтез T³He, в этом случае, выглядит как взаимодействие ³He со связанной системой n-d, а также взаимодействие дейтрона d (из ядра 3He) с тритоном T, в результате чего открываются термоядерные каналы $d\,^3\text{He} \to p\,^4\text{He}$, $dT \to n\,^4\text{He}$, а взаимодействие нейтрона с ядром гелия-3 и нейтрона с тритоном в данном конкретном канале T(³He, np)⁴He остаётся упругим. Реакция синтеза T(³He, np)⁴He с кластерным представлением как ядра-мишени, так и налетающего ядра представляет уникальную возможность тестирования точных методов динамики трёх тел с корректной трактовкой двух- и трёх-частичных каналов с изменением сортов частиц. Отметим, что помимо упругого T³He рассеяния экспериментально известны [3,4,5,6,7] три термоядерных канала

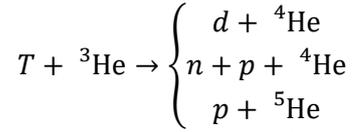

$$T + {}^3\text{He} \to \begin{cases} d + {}^4\text{He} \\ n + p + {}^4\text{He} \\ p + {}^5\text{He} \end{cases}$$

Данные о теоретически возможном канале $T + {}^3\text{He} \to n + {}^5\text{Li}$, а также о вкладе возбужденных состояний промежуточных $^5$He и $^5$Li ядер отсутствуют. Однако, даже те не многочисленные данные по полным сечениям процессов синтеза T$^3$He отличаются в разы по суммарному сечению реакций [4,5], и недостаточно подробно отражают вклад нейтронного канала T($^3$He, np)$^4$He [4,6]. Современные экспериментальные работы 2021 [8] и 2024 [9] годов, как и ранее работа [7], были сосредоточены на угловом распределении протонов и дейтронов в процессах T($^3$He, p)$^5$He и T($^3$He, d)$^4$He без оценивания полного сечения. Учитывая что опыты с $\beta$-радиоактивным, химически активным и высоко подвижным тритием уже несколько десятилетий практически не проводятся, то остаются только теоретические инструменты оценки сечений реакций с ним.

В данной работе продемонстрирована возможность расчёта сечений процессов T($^3$He, np)$^4$He с опорой на интегральные уравнения Фаддеева, описывающие взаимодействие трёх тел с переходом в двух и трёхчастичные (развальные по отношению к кластерному ядру) каналы. Безусловный плюс такого подхода заключается в том, что все развальные каналы получаются автоматически из системы решаемых уравнений для трёхчастичных T-матриц, тогда как каналы с двумя частицами соответствуют упругому рассеянию и реакциям с изменением типа частиц и также могут быть получены из той же системы уравнений. Другими словами трёхчастичный канал не рассматривается отдельно от двухчастичных, а неотделим от него, что отражается на унитарном виде трёхчастичной матрицы рассеяния. Алгоритм расчёта сечения процесса T($^3$He, np)$^4$He был следующим: выписана система уравнений Фаддеева для трёх тел с изменением типа частиц при переходе от одного представления $\alpha \to \beta$ к другому, заданы парные двухчастичные матрицы рассеяния: $t_{np}$ [10] – сепарабельная Боннская модель, $t_{n^3\text{He}}, t_{n^4\text{He}}$ [11] – параметризации по фазовым сдвигам, $t_{nd}$ [12] – микроскопическая трёхчастичная модель, $t_{d^3\text{He}}, t_{dT}$ [11] – потенциальная модель с фактором Гамова. Отличие между $p^4$He и $n^4$He взаимодействиями заключалось в простом добавлении к короткодействующему потенциалу экранированного кулоновского потенциала. Проведённые расчёты с фактором Гамова показали отличие от имеющихся в литературе оценённых ядерных данных ENDF [13]. В работе также проведены оценки четырёхчастичных сечений процессов ионизации с реакцией (y⁻T)($^3$He, npy⁻)$^4$He с атомом-мишенью y⁻T в Борновском приближении с найденной на предыдущем этапе трёхчастичной матрицей реакции T($^3$He, np)$^4$He. Показано, что замена массы электрона на массу мюона характеризуется более объёмным четырёхчастичным фазовым пространством, который формирует максимум в области $E \approx 200$ кэВ при учёте собственных волновых функций системы y⁻T, и практически не меняется в первом Борновском приближении для точной четырёхчастичной матрицы рассеяния.

## 1. Модели парных взаимодействий

Взаимодействие налетающего ядра $^3$He с тритоном, выступающим в роли мишени, который в кластерном представлении является связанной системой нейтрон-дейтрон, требует определения целого ряда промежуточных $t$−матриц рассеяния. Следующая ниже схема показывает как зафиксированы за конкретными ядрами в работе индекс разбиения $i \in [1,2,3]$, выделяющий частицу-спектатор, и индексы $\alpha, \beta$-представления каналов, в которых может находиться система

$$\begin{matrix} & 1 & 2 & 3 \\ \alpha & {}^3\text{He} & n & d \\ \beta & p & n & {}^4\text{He} \end{matrix} : \quad {}^3\text{He} + (nd) \to n + p + {}^4\text{He} \qquad (1.1)$$

Аналогичным образом получается схема для кластерного представления налетающего ядра-гелия-3 путём замены в начальном канале $\alpha$ гелия-3 на тритон, и в обоих каналах нейтрон на протон.

Отличительной чертой проводимых расчётов является тот факт, что уже на этапе подготовки двухчастичных $t-$матриц, соответствующих схеме (1.1) использовалась трёхтельная система уравнений Фаддеева для поиска амплитуд упругого $nd$ рассеяния с Боннским NN потенциалом [10]. Расчёты, проведённые в работе [12] показали, что для описания упругого $nd$ рассеяния достаточно использовать Паде-аппроксимант вида [2/2] с последующей записью найденных амплитуд $U(\boldsymbol{q}_0, \boldsymbol{q})$ в файл в виде числовых таблиц, содержащих саму амплитуду, модули относительных импульсов до $\boldsymbol{q}_0$ и после $\boldsymbol{q}$ рассеяния, а также полярный и азимутальный угол между ними. При расчётах реакции синтеза T($^3$He, np)$^4$He амплитуды $U(\boldsymbol{q}_0, \boldsymbol{q})$ упругого $nd$ рассеяния использовались из ранее записанных в общую память массивов с критерием выбора значения по ближайшей точке. Различий между $pd$ и $nd$ взаимодействиями в данных расчётах не делалось. Взаимодействие нейтронов с ядрами гелия-3 и гелий-4 известно достаточно хорошо для низших парциальных волн, что и определило использование [11] именно сепарабельной параметризации соответствующих парциальных потенциалов вида

$$V_l^J(p_f, p_i) = \lambda_l^J \zeta_l(p_f) \zeta_l(p_i), \quad \zeta_l(p) = \frac{c_1 p^l}{(p^2 + b_1^2)^{l+1}} + \frac{c_2 p^{l+2}}{(p^2 + b_2^2)^{l+2}} \qquad (1.2)$$

с последующим подбором свободных констант $c_1, c_2, b_1,$ и $b_2$ по доступным в литературе данным о парциальных фазовых сдвигах. Взаимодействия в подсистемах $pT$ и $n^3$He для простоты считалось тождественным. Параметры потенциалов (1.2) для $n^3$He и $n^4$He систем приведены в Таблице (1) для фиксированных полных угловых моментов $J$ спина $s$ и орбитальной волны $l$.

Таблица 1. Параметры короткодействующих сепарабельных потенциалов (1.2) для упругих $n^4$He и $n^3$He взаимодействий в соответствующих низших парциальных волнах.

| $^{2s+1}l_J$ | $c_1$, [МэВ]$^{l+\frac{3}{2}}$ | $b_1$, [МэВ] | $c_2$, [МэВ]$^{l+\frac{3}{2}}$ | $b_2$, [МэВ] | $\lambda^{-1}$, [МэВ] |
|---|---|---|---|---|---|
| $^1$S$_{1/2}$ | $14.1 \cdot 10^3$ | 150.1 | 3060 | 124.1 | $W - 4668.7$ |
| $^2$P$_{1/2}$ | $5.88 \cdot 10^5$ | 125.1 | 790 | 1.3 | $-10^{-4}$ |
| $^2$P$_{3/2}$ | 85 | 106.1 | 4020 | 200 | $-10^{-4}$ |
| $^2$D$_{3/2}$ | 110 | 48 | $2.34 \cdot 10^5$ | 173 | $-10^{-4}$ |
| $^1$S$_0$ | 40 | 16.1 | 315.1 | 34.1 | 5 |
| $^3$S$_1$ | 47.1 | 184 | 31.1 | 57.1 | $2 \cdot 10^{-2}$ |
| $^1$P$_1$ | $10^3$ | 1.1 | $4.36 \cdot 10^5$ | 67.1 | $1 \cdot 10^{-2}$ |
| $^3$P$_0$ | 18 | 1.1 | $7.5 \cdot 10^3$ | 72.1 | $6 \cdot 10^{-2}$ |
| $^3$P$_1$ | 46 | 2.3 | 8010 | 68 | $7 \cdot 10^{-2}$ |
| $^3$P$_2$ | 43 | 2.3 | 7830 | 70 | $7 \cdot 10^{-2}$ |
| $^1$D$_2$ | 2.1 | 1.1 | $1.12 \cdot 10^6$ | 118 | $2 \cdot 10^{-2}$ |

Взаимодействие в подсистемах $d^3$He и $dT$ восстанавливалось в соответствии с феноменологическим подходом работы [11] с тремя моделями, описывающими связанные каналы упругого рассеяния и термоядерный канал. Известно, что уравнения Фаддеева многократно проецируют двухчастичные $t-$матрицы на те области фазового пространства, которые не совпадают с областью переменных, для которых эта матрица параметризовывалась с опорой только на двухчастичные экспериментальные наблюдаемые. По этой причине зависимость трёхчастичных расчётов от внемассового поведения двухчастичных $t-$матриц, анализировалась на основе сравнения трёх различных моделей, описывающих подсистемы $d^3$He и $dT$. В модели изолированного резонанса (Res.-модель) простая параметризация в форме Брейта-Вигнера задаёт амплитуду рассеяния с парциальными ширинами, зависящими от конечного относительного импульса. В потенциальной модели (Pot.-модель) локальный потенциал в виде Фурье образа потенциала Юкаве, имея всего два свободных параметра, позволяет описать форму сечения в одном порядке энергий. Наконец, в модели с аналитическим продолжением решаемых уравнений Липпмана-Швингера (A-Cont.-модель) на нефизический лист энергий, где расположены полюса матрицы рассеяния, соответствующие промежуточным резонансам в реакциях синтеза $d^3\text{He} \to p^4\text{He}$ и $dT \to n^4\text{He}$, функциональная зависимость потенциала Pot.-модели меняется согласно работе [14]. Сравнение моделей Pot. и A-Cont. позволяет оценить переход с первого (физического) листа энергий на второй (нефизический) лист, на котором расположены резонансные особенности матрицы рассеяния.

В низкоэнергетических реакциях между ядрами кулоновское отталкивание становится решающим фактором, нивелирующим скорость синтеза. Прямое решение систем уравнений Липпмана-Швингера с кулоновским потенциалом, ввиду кулоновских сингулярностей невозможно. В импульсном пространстве состояния рассеяния не нормированы на единицу и понимаются в обобщённом смысле, что в итоге приводит к тому, что кулоновское отталкивание проявляется, прежде всего, в резольвенте короткодействующей матрицы рассеяния, снижая вероятность рассеяния. Прямой учёт этого эффекта, помимо простого добавления кулоновской $t-$матрицы $T^C$ к короткодействующей $t-$матрице, должен приводить к фактору затухания Гамова. В данной работе отталкивание положительно заряженных ядер при низких энергий учитывается на основе аналитического решения уравнения Шрёдингера для чисто кулоновской задачи рассеяния в трёхмерном пространстве в координатном представлении

$$\Psi_k(r) = e^{ikr} \cdot e^{-\frac{\pi\eta}{2}} \cdot \Gamma(1+i\eta) \cdot F(-i\eta; 1; ikr(1-\cos(\theta))) \quad (1.3)$$

Здесь в формуле (1.3) введены обозначения $\eta$ для параметра Зоммерфельда, $k$ для разницы импульсов до и после рассеяния и $\theta$ для угла рассеяния. Гипергеометрическая функция $F$ в правой части (1.3) для рассеяния на малых углах $\cos(\theta) \to 1$ сама стремится к единице, что снимает зависимость от радиальной координаты $r$ и приводит к простой формуле

$$|\Psi_k(r)|^2 = \frac{2\pi\eta}{e^{2\pi\eta}-1}, \quad (1.4)$$

позволяющей учесть отталкивание на расстояниях $r \approx 0$. Формула (1.4) переходит к формулу Гамова для $\eta \to \infty$. В расчётах кулоновского отталкивания ядер непосредственно используется квадрат функции (1.3) с радиусом $r = 10$ фм, отражающий угловую зависимость, и формула (1.4). Поскольку в кинематике трёх тел относительные импульсы до и после рассеяния не связаны однозначно с двухчастичной энергией рассеяния как в задаче двух тел, то для корректного воспроизведения низкоэнергетического поведения реакции T($^3$He, np)$^4$He сечение корректировалось квадратом волновой функции (1.3) и фактором (1.4) независимо от короткодействующих моделей $d^3\text{He}$ и $dT$ взаимодействий. Последние сами включали аналогичные факторы затухания в явном виде.

## 2. Кинематика процессов

При изучении процессов вида

$$1(E_1, \boldsymbol{q}_0) + (23)(W-E_1, -\boldsymbol{q}_0) \to 1'(E_1', \boldsymbol{q}) + 2'(E_2', \boldsymbol{p}_2') + 3'(W-E_1'-E_2', \boldsymbol{p}_3') \quad (2.1)$$

при некоторой фиксированной полной энергии $W$, относительном импульсе налетающей частицы $\boldsymbol{q}_0$ важно учесть, что подсистема (23) находится в связанном (атомном) состоянии, а вылет частиц $1', 2'$, и $3'$ сопровождается тепловым эффектом $Q > 0$ (для термоядерных реакций). Энергии и импульсы частиц в (2.1) приведены в скобках. Таким образом, для массы $M_t$ системы (23) и кинетической энергии $T$ налетающей частицы (в лабораторной системе) инвариантная масса подсистем $2'3'$ меняется в диапазоне

$$\omega_{23}' \in \left[m_2' + m_3' + Q, m_1 + M_t + T \cdot \frac{M_t}{M_t + m_1} - m_1'\right]. \quad (2.2)$$

Задание инвариантных масс (2.2) позволяет не только верно находить энергию $E_1'$ частицы $1'$, но и относительный импульс $p_{23}'$ любой из частиц в подсистеме $2'3'$

$$E_1' = \frac{W^2 + m_1'^2 - \omega_{23}'^2}{2W}, \quad p_{23}' = \frac{\sqrt{\left(\omega_{23}'^2 - (m_2' - m_3')^2\right)\left(\omega_{23}'^2 - (m_2' + m_3')^2\right)}}{2\omega_{23}'} \quad (2.3)$$

Величины (2.3) необходимы для осуществления Лоренц-преобразования из подсистемы $2'3'$ в общую систему центра масс. Импульсы $\boldsymbol{p}_2'$ и $\boldsymbol{p}_3'$ частиц $2'$ и $3'$ в общей системе центра масс получаются из преобразования Лоренца

$$\boldsymbol{p}_{2',3'} = \pm \boldsymbol{p}_{23}' + \left(-\frac{\boldsymbol{q}}{\omega_{23}'}\right)\left[\frac{\mp \boldsymbol{p}_{23}'\boldsymbol{q}}{W - E_1' + \omega_{23}'} + \frac{\omega_{23}'}{2}\right] \tag{2.4}$$

Частицам $2'$ и $3'$ в (2.4) соответствуют верхний и нижний знаки. Корректное восстановление импульсов всех конечных частиц, вылетающих в непрерывном спектре энергий необходимо для поиска Лоренц-инвариантного сечения процесса (2.1)

$$\frac{d^3\sigma}{d\omega_{23}'d\Omega_{23}'d\Omega_1'} = \frac{E_1(W-E_1)qp_{23}'E_1'E_2'E_3'}{q_0 W^2 (2\pi)^5} \frac{\sum |U_0(\boldsymbol{p},\boldsymbol{q};\boldsymbol{q}_0)|^2}{(2j_1+1)(2j_2+1)} \tag{2.5}$$

Суммирование квадратов амплитуды $U_0(\boldsymbol{p},\boldsymbol{q};\boldsymbol{q}_0)$ развала системы осуществляется по проекциям спинов конечных частиц. Для потенциалов, не содержащих операторов, действующих в спиновом пространстве суммирование в (2.5) можно провести явно. Спины сталкивающихся ядер в (2.5) обозначены как $j_1$ и $j_2$. Из точной формулы для сечения (2.5) процесса (2.1) видно, что энергии конечных частиц $E_1', E_2',$ и $E_3'$ входят в числитель формулы. Это означает, что для одной и той же энергии $W$ и при фиксированных телесных углах вылета $\Omega_1'$ и $\Omega_{23}'$ частицы $1'$ и частиц $2'$ и $3'$ в их собственной подсистеме простое увеличение массы любой из конечных частиц увеличивает фазовый объём реакции (2.1).

Для процессов с четырьмя частицами в конечном состоянии

$$1(E_1, \boldsymbol{q}_0) + (23)(W - E_1, -\boldsymbol{q}_0) \to 1'(E_1', \boldsymbol{q}) + 2'(E_2', \boldsymbol{p}_2') + 3'(E_3', \boldsymbol{p}_3') + 4'(E_4', \boldsymbol{p}_4'), \tag{2.6}$$

которые имеют место в реакциях $^3$He+(Ty$^-$)→n+p+$^4$He+y$^-$ с ионизацией атома Ty$^-$, где y$^-$∈[e$^-$,μ$^-$], полное сечение также пропорционально энергиям $E_1', E_2', E_3',$ и $E_4'$ конечных частиц $1', 2', 3',$ и $4'$, соответственно. Этому обстоятельству, которое во многом определяет различие во вкладах в сечение ионизации электронного и мюонного атомов, уделяется недостаточно внимания в литературе. Для массы $M_a$ атома (23) и кинетической энергии $T$ налетающей частицы (в лабораторной системе) инвариантные массы $\omega_{12}'$ и $\omega_{34}'$ подсистем, для определённости, $1'2'$ и $3'4'$, меняются в диапазонах

$$\omega_{12}' \in \left[m_1' + m_2' + Q, m_1 + M_a + T \cdot \frac{M_a}{M_a + m_1} - m_3' - m_4'\right], \tag{2.7}$$

$$\omega_{34}' \in [m_3' + m_4', W - \omega_{12}']$$

Модули относительных импульсов $p_{12}'$ и $p_{34}'$ любой из частиц в каждой из выделенных подсистем определяются по формуле (2.3) по своим инвариантным массам. Модуль импульса $P$ с которым одна пара конечных частиц движется относительно другой пары конечных частиц также может быть получен из законов сохранения энергии и импульса в виде

$$P = \frac{\sqrt{(W^2 - {\omega_{12}'}^2 - {\omega_{34}'}^2)^2 - 4{\omega_{12}'}^2{\omega_{34}'}^2}}{2W} \tag{2.8}$$

Лоренц-преобразования в общую систему центра масс из каждой подсистемы частиц принимают вид

$$\boldsymbol{p}_{1',2'} = \pm \boldsymbol{p}_{12}' - \left(-\frac{\boldsymbol{P}}{\omega_{12}'}\right)\left[\frac{\pm \boldsymbol{p}_{12}'\boldsymbol{P}}{\sqrt{{\omega_{12}'}^2 + P^2} + \omega_{12}'} + \frac{\omega_{12}'}{2}\right] \tag{2.9}$$

$$\boldsymbol{p}_{3',4'} = \pm \boldsymbol{p}_{12}' + \left(-\frac{\boldsymbol{P}}{\omega_{12}'}\right)\left[\frac{\mp \boldsymbol{p}_{34}'\boldsymbol{P}}{W - \omega_{12}' + \omega_{34}'} + \frac{\omega_{34}'}{2}\right]$$

По импульсам (2.9) можно найти энергии соответствующих частиц, размещая по одной из частиц в паре на массовой поверхности, т.е. принимая условие $E^2 = p^2 + m^2$, а для другой частицы в паре рассчитывая энергию как разницу энергии всей подсистемы и энергии частицы, находящейся на массовой поверхности. По восстановленной кинематике процессов (2.6) можно найти Лоренц-инвариантное выражение для сечения

$$\frac{d^5\sigma}{d\omega'_{12}d\omega'_{34}d\Omega_P d\Omega'_{12}d\Omega'_{34}} = \frac{E_1(W-E_1)p'_{12}p'_{34}E'_1E'_2E'_3E'_4}{q_0 W^2(2\pi)^8} \frac{\sum|U_0(\boldsymbol{p},\boldsymbol{q},\boldsymbol{s};\boldsymbol{q}_0)|^2}{(2j_1+1)(2j_2+1)} \quad (2.10)$$

Выражения для сечений (2.5) и (2.10) являются точными выражениями, в которых нормировка волновых функций начальных и конечных частиц не зависит от энергии и массы частиц. Для получения аналогичного релятивистского выражения необходимо восстановить энергетические множители от нормировок волновых функций, что приведёт к изменению размерности развальных матриц рассеяния $U_0(\boldsymbol{p},\boldsymbol{q};\boldsymbol{q}_0)$ и $U_0(\boldsymbol{p},\boldsymbol{q},\boldsymbol{s};\boldsymbol{q}_0)$, а также потребует для их нахождения решение уравнений типа Бете-Солпитера и его малочастичных аналогов, которые к настоящему моменту плохо изучены. Отметим, что азимутальная симметрия процессов (2.1) и (2.6) сводится к тому, что соответствующие телесные углы $\Omega'_1$ и $\Omega_P$ содержат не тривиальную зависимость только от полярного угла, отсчитываемого от направления $\boldsymbol{q_0}$-импульса налетающей частицы. На Рис. 1 приведено отношение сечений, рассчитываемых по формуле (2.10) для термоядерной реакции (y⁻T)(³He, npy⁻)⁴He с ионизацией атома y⁻T, для случаев когда y⁻ – это мюон и когда y⁻ – это электрон. Пунктирная кривая демонстрирует роль различных объёмов фазовых пространств, связанных только с различием в массах лептонов при той же полной энергии $W$ сталкивающихся частиц. Видно, что в случае с мюонным атомом без учёта динамических особенностей реакций сечение процесса в несколько тысяч раз должно быть больше уже при малых относительных энергиях. Добавление собственных волновых функций для мюонного и электронного атома [15], которые как показано следующем разделе входят в неоднородное слагаемое трёхчастичных динамических уравнений Фаддеева, меняет отношение фазовых объёмов. В этом случае роль мюона становится ещё более выраженной по сравнению с ролью электрона.

Рис. 1. Отношение сечений процессов (y⁻T)(³He, npy⁻)⁴He для y⁻=μ⁻ и y⁻=e⁻, рассчитываемых по формуле (2.10). Пунктирная линия – отношение только фазовых множителей, точечная линия – дополнительный учёт собственных волновых функций начального атома в $l=0$ парциальной волне, пустые точки – дополнительный учёт трухчастичных матриц рассеяния, соответствующий первому Борновскому приближению к четырёхчастичной задаче рассеяния.

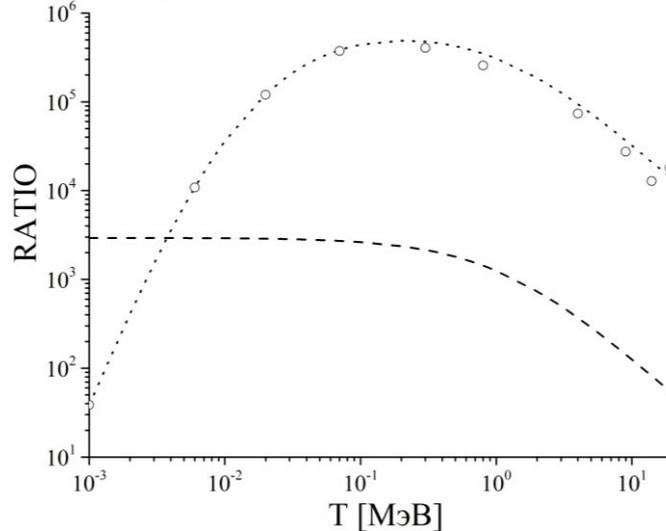

Следует отметить, что конкурирующий с термоядерным канал развала начального ядра-мишени или атома на составные частицы является эндотермическим процессом, который тем не менее, как будет видно из следующего раздела, присутствует в динамике нескольких тел. Для этого канала кинематические соотношения (2.2-2.4) и (2.7-2.9) остаются справедливыми, принимая $Q=0$, а сам канал отрывается по достижению $T$ пороговой энергии.

## 3. Динамика процесса T($^3$He, np)$^4$He в задаче трёх тел

Для нахождения сечения (2.5) необходимо связать развальную амплитуду $U_0(\boldsymbol{p},\boldsymbol{q};\boldsymbol{q}_0)$ с трёхчастичными $T$ −матрицами рассеяния. Поскольку индекс разбиения $i$ развальных матриц привязан к вылетающей из системы частицы, то можно сразу в представлении импульсов $\boldsymbol{p}$ и $\boldsymbol{q}$ записать

$$U_0(\boldsymbol{p},\boldsymbol{q};\boldsymbol{q}_0) = \langle \boldsymbol{pq}|T_1\Psi_{1(23)}\rangle + \langle \widetilde{\boldsymbol{pq}}|T_2\Psi_{2(31)}\rangle + \langle \overline{\boldsymbol{pq}}|T_3\Psi_{3(12)}\rangle \quad (3.1)$$

Импульсы $\boldsymbol{p}$ и $\boldsymbol{q}$ в (3.1) ассоциированы с соответствующими наборами переменных Якоби и отвечают относительному импульсу в выделенной паре частиц и импульсу третьей частицы относительно этой пары. В работе [12] показано, что пары импульсов $\widetilde{\boldsymbol{pq}}$ и $\overline{\boldsymbol{pq}}$ необходимые для расчёта развальных матриц $T_2$ и $T_3$, могут быть получены действием операторов перестановок номеров частиц на выражения для импульсов $\boldsymbol{p}$ и $\boldsymbol{q}$. Однако, непосредственным сравнением можно убедиться, что эти же импульсы можно получить из общей системы центра масс, определяя все три набора переменных Якоби. Собственные функции $\Psi_{i(jk)}$ в (3.1) определяют выделенную подсистему $(jk)$ в начальном состоянии, которая может характеризоваться определённой энергией связи и собственной волновой функцией. Также выражение $\Psi_{i(jk)}$ (где $i \ne j \ne k$) может включать, в общем случае, не только плосковолновое начальное состояние третьей частицы $i$, но и волновой пакет.

С точки зрения динамики трёх тел и общей унитарности трёхчастичной матрицы рассеяния процессы с двумя и тремя частицами в конечном состоянии являются связанными. Это означает, что операторы перестройки системы, описывающие переход из состояний с двумя частицами в состояния с двумя другими частицами, удовлетворяющие своей собственной системе зацепляющихся интегральных уравнений, могут быть получены с помощью развальных трёхчастичных $T$ −матриц рассеяния простыми квадратурными формулами. В данной работе расчёты ограничены только каналом с тремя частицами в конечном состоянии, поэтому выражение (3.1) для амплитуды является достаточным.

Уравнение Фаддеева для трёхчастичной развальной $T$ −матрицы получается естественным расширением векторной формы уравнений [12] на пространство каналов (1.1) по правилу $T_i \to T_i^{\alpha\beta}$

$$\begin{pmatrix} T_1^{11} \\ T_1^{21} \\ T_2^{11} \\ T_2^{21} \\ T_3^{11} \\ T_3^{21} \end{pmatrix} = \begin{pmatrix} t_1^{11}(\phi_2+\phi_3) \\ t_1^{21}(\phi_2+\phi_3) \\ t_2^{11}(\phi_3+\phi_1) \\ t_2^{21}(\phi_3+\phi_1) \\ t_3^{11}(\phi_1+\phi_2) \\ t_3^{21}(\phi_1+\phi_2) \end{pmatrix} + \begin{pmatrix} 0 & 0 & t_1^{11}G_0 & 0 & t_1^{11}G_0 & 0 \\ 0 & 0 & 0 & t_1^{22}G_0 & 0 & t_1^{22}G_0 \\ t_2^{11}G_0 & t_2^{12}G_0 & 0 & 0 & t_2^{11}G_0 & t_2^{12}G_0 \\ t_2^{21}G_0 & t_2^{22}G_0 & 0 & 0 & t_2^{21}G_0 & t_2^{22}G_0 \\ t_3^{11}G_0 & 0 & t_3^{11}G_0 & 0 & 0 & 0 \\ 0 & t_3^{22}G_0 & 0 & t_3^{22}G_0 & 0 & 0 \end{pmatrix} \begin{pmatrix} T_1^{11} \\ T_1^{21} \\ T_2^{11} \\ T_2^{21} \\ T_3^{11} \\ T_3^{21} \end{pmatrix}, \quad (3.2)$$

где двухчастичные $t$ −матрицы относятся к разным пространствам представлений. Одинаковый верхний индекс отвечает процессам без изменения типов частиц, тогда как изменение пространства представлений $1 \to 2$ отвечает изменению типов частиц, т.е. рассматриваемой термоядерной реакции (1.1). Отсюда видно, что для рассматриваемой задачи представляет интерес нахождение величин $T_i^{21}$, с $i = 1,2,3$. Система (3.2) имеет симметричный, относительно перестановки индексов частиц, вид и для конкретного процесса (1.1), имея всего одно связанное $nd$ состояние, имеем $\phi_2 = 0$ и $\phi_3 = 0$. Кроме того, в подсистемах с номерами (12) и (23) отсутствуют по определению ядерные реакции, поэтому соответствующие двухчастичные $t$ −матрицы в пространстве каналов имеют диагональный вид. Двухчастичные функции Грина $G_0$ в (3.2) имеют энергетический сдвиг, связанный с движением третьей частицы, что качественно отличает динамику двух тел и движение частицы в эффективном поле от точной динамической картины. Этот сдвиг, разный для каждого набора переменных Якоби, задаёт энергию рассеяния $z$ в двухчастичной подсистеме

$$z = E - \frac{q_{i(jk)}^2}{2M_{i(jk)}}, \qquad E = T\frac{M_t}{M_t + m_1} - |E_b| \quad (3.3)$$

Энергия рассеяния трёх тел $E$, как видно из формулы (3.3), связана с кинетической энергией налетающей частицы и отражает зависимость от энергии связи подсистем (23) процессов (2.1). Из (3.3) также видно, что величины $E$ и $z$ могут быть как положительными так и отрицательными, что меняет распределение на комплексной плоскости двухчастичных полюсов, которые генерируют структуру двухчастичных $t$−матриц. Присутствие третьей частицы видоизменяет область разреза резольвент в т.н. луноподобные области, известные как области логарифмических сингулярностей. Записывая любую из резольвент $G_0$ правой части (3.2) в виде $G_0 \equiv m/(qq'')/(y_0 - x'')$, где $q''$-радиальная переменная интегрирования, а $x''$-косинус полярного угла интегрирования в (3.2), для любой из итераций уравнения (3.2) можно получить выражение, явно содержащее логарифмическую сингулярность

$$\int q''^2 dq'' d\Omega_{q''} t \frac{m}{qq''} \frac{t}{y_0 - x''} \equiv \int q'' dq'' \int_{-1}^{1} dx'' \frac{f(q'', x'')}{y_0 - x'' + i\epsilon} =$$

$$= \int q'' dq'' \left[ \int_{-1}^{1} dx'' \frac{f(q'', x'') - \hat{f}(q'', y_0)}{y_0 - x''} + \int_{-1}^{1} dx'' \frac{\hat{f}(q'', y_0)}{y_0 - x''} \right] \equiv \qquad (3.4)$$

$$\int q'' dq'' \left[ \int_{-1}^{1} dx'' \frac{f(q'', x'') - \hat{f}(q'', y_0)}{y_0 - x''} + \hat{f}(q'', y_0) \ln\left(\left|\frac{1 + y_0}{1 - y_0}\right|\right) \right.$$

$$\left. - \Theta(1 - |y_0|) i\pi \hat{f}(q'', y_0) \right].$$

Аргументы и индексы у двухчастичных $t$−матриц в левой части (3.4) для простоты опущены. Величина $\hat{f}(q'', y_0)$ равна исходной функции $f(q'', y_0)$ когда $|y_0| \leq 1$ и равна $f(q'', y_0/|y_0|)$ когда $|y_0| > 1$. Функция Хевисайда $\Theta$ в (3.4) контролирует попадание в круг $|y_0| \leq 1$, при котором в интеграле (3.4) появляется полюсная добавка. Отметим, что последние два слагаемых в нижней строчке (3.4) появляются в непересекающихся кинематических ситуациях. Пересечение областей логарифмических сингулярностей при интегрировании удобнее производить с помощью интерполирования сплайном, контролируя границы этой области.

Присутствие каналов с перестройкой в системе трёх тел приводит к тому, что в области $E > 0$ алгебраическая аппроксимация однородной формы интегрального уравнения (3.2) имеет нетривиальные решения (т.е. связанные состояния), из-за чего обращение с уравнением (3.2) как с матричным на численной сетке импульсов, не позволяет путём обращения матриц находить решение системы. Поэтому решение системы (3.2) ищется в виде отдельных итераций

$$T_i^{\alpha\beta} \approx \sum_n (C_n)_i^{\alpha\beta}, \qquad (3.5)$$

где $n = 0$ соответствует неоднородному слагаемому системы (3.2). Каждая итерация в (3.5) удовлетворяет системе рекуррентных зацепляющихся соотношений

$$(C_n)_1^{11} = t_1^{11} G_0 (C_{n-1})_2^{11} + t_1^{11} G_0 (C_{n-1})_3^{11}$$
$$(C_n)_1^{21} = t_1^{22} G_0 (C_{n-1})_2^{21} + t_1^{22} G_0 (C_{n-1})_3^{21}$$
$$(C_n)_2^{11} = t_2^{11} G_0 [(C_{n-1})_1^{11} + (C_{n-1})_3^{11}] + t_2^{12} G_0 [(C_{n-1})_1^{21} + (C_{n-1})_3^{21}]$$
$$(C_n)_2^{21} = t_2^{21} G_0 [(C_{n-1})_1^{11} + (C_{n-1})_3^{11}] + t_2^{22} G_0 [(C_{n-1})_1^{21} + (C_{n-1})_3^{21}] \qquad (3.6)$$
$$(C_n)_3^{11} = t_3^{11} G_0 (C_{n-1})_1^{11} + t_3^{11} G_0 (C_{n-1})_2^{11}$$
$$(C_n)_3^{21} = t_3^{22} G_0 (C_{n-1})_1^{21} + t_3^{21} G_0 (C_{n-1})_2^{21}.$$

Прямое решение (3.6) было выполнено до $n = 2$.

## 4. Результаты и заключение

Полное сечение процесса $T + {}^3\text{He} \to n + p + {}^4\text{He}$, рассчитанное по формуле (2.5) с амплитудой, получаемой из уравнений (3.1), (3.5) и (3.6) приведено на Рис.2. Расчёты выполнены с тремя моделями $d^3\text{He}$ взаимодействий (Res.-точечная кривая, Pot.-штрих-пунктирная кривая, A-Cont.-слошная кривая). Добавление кластерного канала $pd$ для налетающего ядра $n^3\text{He}$, которое приводит к тому, что конечное $n^4\text{He}$ состояние формируется за счёт $dt$ реакции синтеза отражено штрих-двойным пунктиром. Без учёта факторов подавления сечения за счёт кулоновского отталкивания ядер расчётные кривые не отражают характерного экспериментального затухания сечений. Учёт отталкивания ядер производился путём домножения квадрата состояния рассеяния (1.3) с фиксированным радиусом $r = 10$ фм на дифференциальное сечение (2.5), с одной стороны, и путём домножения полного сечения на фактор (1.4), с другой стороны. В первом случае удаётся отследить угловую зависимость фактора подавления и неточечный характер столкновения, которые приводят к тому, что итоговое сечение получается на порядок выше чем результат простого домножения на фактор (1.4). Влияние факторов подавления сечений, рассчитываемых с использованием Pot.-модели $d^3\text{He}$ и $dt$ взаимодействий отражено затенённой областью на Рис.2. Видно, что итоговое сечение в области $T < 100$ кэВ где нет экспериментальных данных выше оценок библиотеки ENDF практически на порядок.

Рис.2. Полное сечение реакции $T + {}^3\text{He} \to n + p + {}^4\text{He}$, полученное в данной работе в сравнении с оценёнными ядерными данными библиотеки ENDF [13] реакций $T + {}^3\text{He} \to n + p + {}^4\text{He}$ и $T + {}^3\text{He} \to d + {}^4\text{He}$. Приведены также известные экспериментальные данные (ромбы, звёзды и кресты [3], треугольники и квадраты [4], круги [5], нижний треугольник [6], пятигранники [7]), относящиеся к полному сечению всех реакций (треугольники и круги), сечению с вылетом нейтронов (квадраты), трёх частиц в непрерывном спектре энергий (звёзды нижний треугольник), протон-гелий-5 каналу (кресты) и дейтрон-гелий-4 каналу (пятигранники). Справа представлена увеличенная область энергий 100-кэВ-20МэВ.

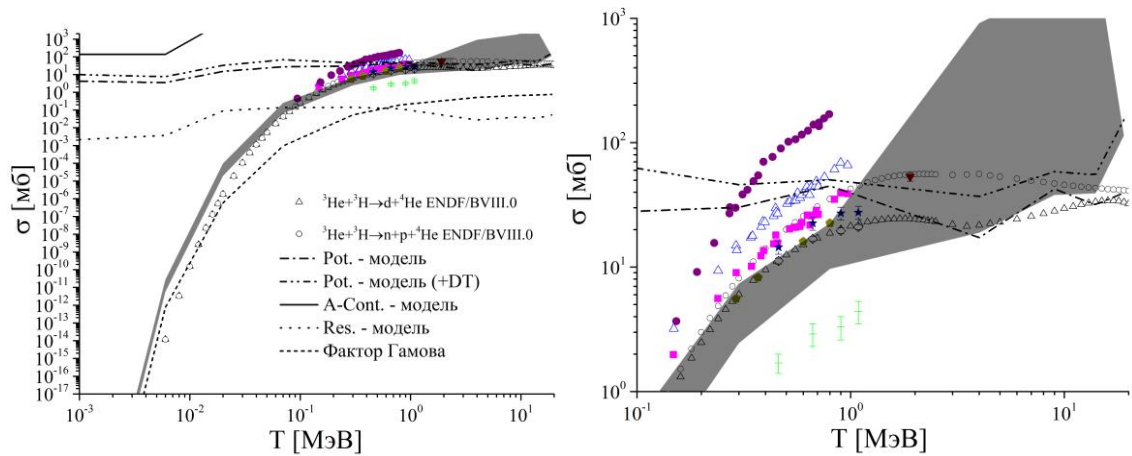

Более широкая часть затенённой области в правой части Рис.2 соответствуют более вероятному рассеянию ядер $^3\text{He}$ и T в задней полусфере углов, в результате чего значения гипергеометрической функции (1.3) могут сильно превышать единицу. Проведённые расчёты показывают отличное согласие с относящимися к реакции $T + {}^3\text{He} \to n + p + {}^4\text{He}$ экспериментальными данными (звёзды), демонстрируя корректность использования кластерного подхода к описанию реакции с использованием точной динамики трёх тел. Полученные оценки могут представлять интерес для кинетических расчётов наработки изотопов в гелий-тритиевой плазме.

В заключительной части работы были проведены оценочные расчёты сечений влияния замены массы электрона на массу мюона в процессе синтеза (y⁻T)(³He, npy⁻)⁴He с ионизацией атома-мишени (y⁻T) где y⁻∈[e⁻,µ⁻]. Амплитуда четырёхчастичного рассеяния $U_0(\boldsymbol{p}, \boldsymbol{q}, \boldsymbol{s}; \boldsymbol{q}_0)$ рассчитывалась в первом Борновском приближении

$$U_0(\boldsymbol{p},\boldsymbol{q},\boldsymbol{s};\boldsymbol{q}_0) \approx U_0(\boldsymbol{p},\boldsymbol{q};\boldsymbol{q}_0)\phi_{\mu^-T}(\boldsymbol{p}) \qquad (4.1)$$

С собственной функцией $\phi_{\mu^-T}(\boldsymbol{p})$ атома ($\mu^-$T). Отношение сечений процесса ($\mu^-$T)($^3$He, np$\mu^-$)$^4$He, полученные с амплитудой (4.1) представлены пустыми точками на Рис.1. Видно, что первое Борновское приближение незначительно отличается от отношения фазовых пространств и собственных волновых функций. Эти расчёты показывают, что замена электрона мюоном в реакции синтеза ($\mu^-$T)($^3$He, np$\mu^-$)$^4$He сильно увеличивает сечение и результат остаётся справедливым в первом Борновском приближении.

## Благодарности

# Расчёт сечений T($^3$He, np)$^4$He и (Ty$^-$)($^3$He, npy$^-$) $^4$He в динамике трёх тел

М.В. Егоров[1,2]

[1]*Физический факультет, Томский государственный университет, г.Томск*

[2]*Лаборатория теоретической физики, Объединённый институт ядерных исследований, г.Дубна*

В работе проведены трёхчастичные расчёты сечений реакции синтеза $^3He+T\rightarrow n+p+^4He$ с помощью решения связанных шести интегральных уравнений Фаддеева с кластерным представлением ядра мишени - тритона как связанной нейтрон-дейтронной системы. Короткодействующие взаимодействия в подсистемах пар частиц включали микроскопические np, nd, n$^4$He, n$^3$He матрицы рассеяния, а также феноменологические d$^3$He модели, допускающие связывание упругого канала с каналом реакции d$^3$He. Кратко обсуждается роль моделей d$^3$He и кулоновского взаимодействия $^3$He и T ядер. Полученная трёхчастичная матрица реакции использована для оценки роли мюона в процессе $^3He+(Ty^-)\rightarrow n+p+^4He+y^-$ с атомом Ty$^-$, где y$^-\in[e^-,\mu^-]$. Показано, что эффект роста сечения вызванный присутствием мюона в зоне реакций, продиктован фазовым множителем и сохраняется в первом Борновском приближении для точной четырёхчастичной матрицы рассеяния.

***Ключевые слова***: *малочастичная динамика, ядерный синтез, гелий-3, тритий, уравнения Фаддеева*

Егоров Михаил Викторович, кандидат физико-математических наук,

- доцент кафедры квантовой теории поля физического факультета национального исследовательского Томского государственного университета,
- научный сотрудник сектора №3 лаборатории теоретической физики Объединённого института ядерных исследований (по совместительству).

Email: egorovphys@mail.ru

# Cross section calculation of T($^3$He, np)$^4$He and (Ty$^-$)($^3$He, npy$^-$) $^4$He in the three-body dynamics

M.V. Egorov[1,2]

$^1$*Physics faculty, Tomsk State University, Tomsk*

$^2$*Bodolyubov Laboratory of Theoretical Physics, Joint Institute for Nuclear Research, Dubna*

*This work presents three-body calculations of the cross sections for the $^3$He+T→n+p+$^4$He fusion reaction. These calculations were performed by solving a set of six coupled Faddeev integral equations, utilizing a cluster representation of the target nucleus, the triton, as a bound neutron-deuteron system. Short-range interactions within the two-body subsystems included microscopic np, nd, n$^4$He, and n$^3$He scattering matrices, as well as phenomenological d$^3$He models that allow for the coupling of the elastic channel with the d$^3$He reaction channel. The roles of the d$^3$He models and the Coulomb interaction between the $^3$He and T nuclei are briefly discussed. The resulting three-body reaction matrix was then utilized to estimate the role of a muon in the process $^3$He+(Ty$^-$)→n+p+$^4$He+y$^-$ involving a Ty$^-$ atom, where y$^-$∈[e$^-$,μ$^-$]. It is demonstrated that the effect of cross-section enhancement induced by the muon's presence in the reaction zone is governed by a phase factor and persists in the first Born approximation for the exact four-body scattering matrix.*

***Keywords****: few-body dynamics, nuclear fusion, helium-3, tritium, Faddeev equations*

Egorov Mikhail Viktorovish, Ph.D. in Physics and Mathematics,

- Associate Professor of the Department of Quantum Field Theory, Faculty of Physics, National Research Tomsk State University,

- Research Fellow, Sector No. 3, Bogolyubov Laboratory of Theoretical Physics, Joint Institute for Nuclear Research (part-time)

Email: egorovphys@mail.ru

Egorov Mikhail gratefully acknowledges the financial support of the Foundation for the Advancement of Theoretical Physics and Mathematics «BASIS» (project No 23-1-3-3-1).

The research was carried out with the support of a grant from the Government of the Russian Federation (Agreement No. 075-15-2025-009 of 28 February 2025).